\documentstyle[12pt]{article}

\catcode`\@=11
\long\def\@makefntext#1{ 
\protect\noindent \hbox to 3.2pt {\hskip-.9pt
$^{{\ninerm\@thefnmark}}$\hfil}#1\hfill} 

\def\thefootnote{\fnsymbol{footnote}}
 \def\@makefnmark{\hbox to 0pt{$^{\@thefnmark}$\hss}}  

\def\ps@myheadings{\let\@mkboth\@gobbletwo
\def\@oddhead{\hbox{} 
\rightmark\hfil\ninerm\thepage}
\def\@oddfoot{}\def\@evenhead{\ninerm\thepage\hfil 
\leftmark\hbox{}}\def\@evenfoot{}
\def\sectionmark##1{}\def\subsectionmark##1{}}

\begin{document}

\newcommand{\symbolfootnote}{\renewcommand{\thefootnote}
	{\fnsymbol{footnote}}}
\renewcommand{\thefootnote}{\fnsymbol{footnote}}
\newcommand{\alphfootnote}
	{\setcounter{footnote}{0}
	 \renewcommand{\thefootnote}{\sevenrm\alph{footnote}}}

\newcounter{sectionc}\newcounter{subsectionc}\newcounter{subsubsectionc}
\renewcommand{\section}[1] {\vspace{0.6cm}\addtocounter{sectionc}{1}
\setcounter{subsectionc}{0}\setcounter{subsubsectionc}{0}\noindent
	{\bf\thesectionc. #1}\par\vspace{0.4cm}}
\renewcommand{\subsection}[1] {\vspace{0.6cm}\addtocounter{subsectionc}{1}
	\setcounter{subsubsectionc}{0}\noindent
	{\it\thesectionc.\thesubsectionc. #1}\par\vspace{0.4cm}}
\renewcommand{\subsubsection}[1]
{\vspace{0.6cm}\addtocounter{subsubsectionc}{1}
	\noindent {\rm\thesectionc.\thesubsectionc.\thesubsubsectionc.
	#1}\par\vspace{0.4cm}}
\newcommand{\nonumsection}[1] {\vspace{0.6cm}\noindent{\bf #1}
	\par\vspace{0.4cm}}

\newcounter{appendixc}
\newcounter{subappendixc}[appendixc]
\newcounter{subsubappendixc}[subappendixc]
\renewcommand{\thesubappendixc}{\Alph{appendixc}.\arabic{subappendixc}}
\renewcommand{\thesubsubappendixc}
	{\Alph{appendixc}.\arabic{subappendixc}.\arabic{subsubappendixc}}

\renewcommand{\appendix}[1] {\vspace{0.6cm}
        \refstepcounter{appendixc}
        \setcounter{figure}{0}
        \setcounter{table}{0}
        \setcounter{equation}{0}
        \renewcommand{\thefigure}{\Alph{appendixc}.\arabic{figure}}
        \renewcommand{\thetable}{\Alph{appendixc}.\arabic{table}}
        \renewcommand{\theappendixc}{\Alph{appendixc}}
        \renewcommand{\theequation}{\Alph{appendixc}.\arabic{equation}}
        \noindent{\bf Appendix \theappendixc #1}\par\vspace{0.4cm}}
\newcommand{\subappendix}[1] {\vspace{0.6cm}
        \refstepcounter{subappendixc}
        \noindent{\bf Appendix \thesubappendixc. #1}\par\vspace{0.4cm}}
\newcommand{\subsubappendix}[1] {\vspace{0.6cm}
        \refstepcounter{subsubappendixc}
        \noindent{\it Appendix \thesubsubappendixc. #1}
	\par\vspace{0.4cm}}

\def\abstracts#1{{
	\centering{\begin{minipage}{30pc}\tenrm\baselineskip=12pt\noindent
	\centerline{\tenrm ABSTRACT}\vspace{0.3cm}
	\parindent=0pt #1
	\end{minipage} }\par}}

\newcommand{\bibit}{\it}
\newcommand{\bibbf}{\bf}
\renewenvironment{thebibliography}[1]
	{\begin{list}{\arabic{enumi}.}
	{\usecounter{enumi}\setlength{\parsep}{0pt}
\setlength{\leftmargin 1.25cm}{\rightmargin 0pt}
	 \setlength{\itemsep}{0pt} \settowidth
	{\labelwidth}{#1.}\sloppy}}{\end{list}}

\topsep=0in\parsep=0in\itemsep=0in
\parindent=1.5pc

\newcounter{itemlistc}
\newcounter{romanlistc}
\newcounter{alphlistc}
\newcounter{arabiclistc}
\newenvironment{itemlist}
    	{\setcounter{itemlistc}{0}
	 \begin{list}{$\bullet$}
	{\usecounter{itemlistc}
	 \setlength{\parsep}{0pt}
	 \setlength{\itemsep}{0pt}}}{\end{list}}

\newenvironment{romanlist}
	{\setcounter{romanlistc}{0}
	 \begin{list}{$($\roman{romanlistc}$)$}
	{\usecounter{romanlistc}
	 \setlength{\parsep}{0pt}
	 \setlength{\itemsep}{0pt}}}{\end{list}}

\newenvironment{alphlist}
	{\setcounter{alphlistc}{0}
	 \begin{list}{$($\alph{alphlistc}$)$}
	{\usecounter{alphlistc}
	 \setlength{\parsep}{0pt}
	 \setlength{\itemsep}{0pt}}}{\end{list}}

\newenvironment{arabiclist}
	{\setcounter{arabiclistc}{0}
	 \begin{list}{\arabic{arabiclistc}}
	{\usecounter{arabiclistc}
	 \setlength{\parsep}{0pt}
	 \setlength{\itemsep}{0pt}}}{\end{list}}

\newcommand{\fcaption}[1]{
        \refstepcounter{figure}
        \setbox\@tempboxa = \hbox{\tenrm Fig.~\thefigure. #1}
        \ifdim \wd\@tempboxa > 6in
           {\begin{center}
        \parbox{6in}{\tenrm\baselineskip=12pt Fig.~\thefigure. #1 }
            \end{center}}
        \else
             {\begin{center}
             {\tenrm Fig.~\thefigure. #1}
              \end{center}}
        \fi}

\newcommand{\tcaption}[1]{
        \refstepcounter{table}
        \setbox\@tempboxa = \hbox{\tenrm Table~\thetable. #1}
        \ifdim \wd\@tempboxa > 6in
           {\begin{center}
        \parbox{6in}{\tenrm\baselineskip=12pt Table~\thetable. #1 }
            \end{center}}
        \else
             {\begin{center}
             {\tenrm Table~\thetable. #1}
              \end{center}}
        \fi}

\def\@citex[#1]#2{\if@filesw\immediate\write\@auxout
	{\string\citation{#2}}\fi
\def\@citea{}\@cite{\@for\@citeb:=#2\do
	{\@citea\def\@citea{,}\@ifundefined
	{b@\@citeb}{{\bf ?}\@warning
	{Citation `\@citeb' on page \thepage \space undefined}}
	{\csname b@\@citeb\endcsname}}}{#1}}

\newif\if@cghi
\def\cite{\@cghitrue\@ifnextchar [{\@tempswatrue
	\@citex}{\@tempswafalse\@citex[]}}
\def\citelow{\@cghifalse\@ifnextchar [{\@tempswatrue
	\@citex}{\@tempswafalse\@citex[]}}
\def\@cite#1#2{{$\null^{#1}$\if@tempswa\typeout
	{IJCGA warning: optional citation argument
	ignored: `#2'} \fi}}
\newcommand{\citeup}{\cite}

\def\fnm#1{$^{\mbox{\scriptsize #1}}$}
\def\fnt#1#2{\footnotetext{\kern-.3em
	{$^{\mbox{\sevenrm #1}}$}{#2}}}

\font\twelvebf=cmbx10 scaled\magstep 1
\font\twelverm=cmr10 scaled\magstep 1
\font\twelveit=cmti10 scaled\magstep 1
\font\elevenbfit=cmbxti10 scaled\magstephalf
\font\elevenbf=cmbx10 scaled\magstephalf
\font\elevenrm=cmr10 scaled\magstephalf
\font\elevenit=cmti10 scaled\magstephalf
\font\bfit=cmbxti10
\font\tenbf=cmbx10
\font\tenrm=cmr10
\font\tenit=cmti10
\font\ninebf=cmbx9
\font\ninerm=cmr9
\font\nineit=cmti9
\font\eightbf=cmbx8
\font\eightrm=cmr8
\font\eightit=cmti8


\hfill ORNL/CCIP/94-15
\vskip 1cm
\centerline{\Large\bf CEBAF AT HIGHER
ENERGIES\footnote{A Workshop held at CEBAF,
Newport News, Virginia, 14-16 April 1994.}}
\baselineskip=22pt
\centerline{\bf WORKING GROUP REPORT ON}
\baselineskip=16pt
\centerline{\bf HADRON SPECTROSCOPY AND PRODUCTION}
\baselineskip=32pt
\vspace{0.3cm}
\centerline{\tenrm TED BARNES}
\baselineskip=13pt
\centerline{\tenit Physics Division, Oak Ridge National Laboratory}
\baselineskip=12pt
\centerline{\tenit Oak Ridge, TN, 37831-6373}
\baselineskip=13pt
\centerline{\tenit and}
\baselineskip=12pt
\centerline{\tenit Department of Physics and Astronomy,
University of Tennessee}
\baselineskip=12pt
\centerline{\tenit Knoxville, TN, 37996-1200}
\vspace{0.3cm}
\centerline{\tenrm and}
\vspace{0.3cm}
\centerline{\tenrm JIM NAPOLITANO}
\baselineskip=13pt
\centerline{\tenit Department of Physics, Rensselaer Polytechnic Institute}
\baselineskip=12pt
\centerline{\tenit Troy, NY, 12180-3590}
\vspace{0.5cm}
\abstracts{This report summarizes topics
in hadron spectroscopy and
production which could be addressed
at CEBAF with an energy upgrade to $E_\gamma=8$ GeV and beyond.
The topics discussed include conventional meson and baryon spectroscopy,
spectroscopy of exotica (especially molecules and hybrids),
CP and CPT tests using
$\phi$ mesons, and new detector and accelerator options.}

\vfil
\rm\baselineskip=14pt
\section{Overview}

The photon, real or virtual, makes a unique particle beam. It has spin one,
and
selectively couples to different quark flavors according to their charges. As
the beam energy increases, one can expect to see a transition from dynamics
dominated by a few hadron resonances to a regime in which perturbative QED
and
QCD processes are evident.\cite{Leith} In the resonance regime, the photon's
polarization allows tests of many resonance form factors which have been
studied theoretically but never adequately investigated experimentally. In
addition to conventional hadronic resonances, unusual states such as
glueballs,
hybrids and molecules are anticipated in the 1-3 GeV region, and these may be
identifiable through unusual photon couplings and decay modes. Obviously, the
opportunities for spectroscopy with a sufficiently energetic and intense
photon
beam are enormous.

Despite these opportunities, experimental progress in resonance production by
photons has been rather limited because high intensity, high energy CW
electron
accelerators have not been available.  The initial program of experiments at
CEBAF with a 4~GeV beam represents the first high-statistics experimental
investigation of this field.  However, most of these experiments are limited
to
baryon spectroscopy in $s$-channel photo- or electroproduction, because the
beam energy is not high enough to reach the thresholds anticipated for
photoproduction of interesting new meson resonances. Of course one should
exceed these expected thresholds in the design beam energy, both for reasons
of
cross section and to insure that states somewhat above theoretical mass
predictions are not missed.

For our quantitative discussion of energies we consider the photoproduction
reaction
$$
\gamma+P\to m+B
$$
where $m$ is a meson (of mass $m$) and $B$ is a baryon (of mass $M_B$).
The photon energy $E_\gamma$ required to reach the threshold for this
reaction
is
$$
E_\gamma=\frac{1}{2M_P}\left[\left(m+M_B\right)^2-M_P^2\right]\ .
\eqno(1)
$$
The highest meson mass is produced against a final-state proton, in which
case
the beam energy required to reach a threshold $m$ is
$$
E_\gamma= m + {m^2\over 2M_P} \ .
\eqno(2)
$$
Electroproduction using timelike photons requires higher photon energy than
photoproduction to reach a given invariant mass. For an off-shell photon with
$Q^2<0$ the threshold energy required for the reaction
$$
\gamma^* + P \to m + P
$$
is given by
$$
E_\gamma^* =  {1\over 2M_P} \left[ \; m^2 + 2mM_P + |Q^2| \; \right] \ .
$$
Thresholds associated with photo- and electroproduction from a proton are
shown as a contour plot in figure 1. As examples, the threshold photon energy
for photoproducing a 1.9 GeV light hybrid is $E_\gamma = 3.8$~GeV, which is
just possible at CEBAF with $E_\gamma=4$ GeV; photoproduction of a 2.5~GeV
$s\bar s$ meson requires $E_\gamma = 5.8$~GeV; the $\psi$ at 3.1 GeV requires
$E_\gamma = 8.2$~GeV; and a charmonium hybrid at 4.1 GeV would require
$E_\gamma = 13.1$~GeV.
\vskip 0.5cm

{\it Thus, with an upgrade of CEBAF to $E_\gamma=8$ GeV one could use
photoproduction to explore meson spectroscopy up to an invariant mass of
about
3.0 GeV but no higher.}

\vskip 0.5cm
This should allow detailed photoproduction and electroproduction studies of
$u,d,s$ $q\bar q$ spectroscopy as well as searches for light hybrids. In
order
to produce the reaction products with sufficient phase space to be useful
experimentally the beam energy should of course be somewhat higher than these
threshold values.

Our working group considered both theoretical and experimental aspects of an
energy upgrade. The preliminary organization was informal, and reserved many
intervals for discussion and additional contributions; this proved to be a
wise
decision, since the talks generated extensive discussions. The topics
considered were generally in the following categories:
\begin{itemize}
\item{Conventional Meson and Baryon Spectroscopy}
\item{Exotica: Molecules and Hybrid Mesons}
\item{$CP$ and $CPT$ Violation in Flying $\phi$ Decay}
\item{New Detectors and Accelerator Developments}
\end{itemize}
We also had two contributions on specialized topics:
Marc Sher described a possible experiment to search for low-mass
gluinos\cite{gluinos} (a proposed spin-1/2, strongly interacting
supersymmetric partner of the gluon) in a mass and lifetime ``window" that
has
not been excluded experimentally.
Kam Seth described the high-precision charmonium results from Fermilab
Experiment E760, and discussed how charm and related
physics could be addressed with a
$\tau$-charm factory at CEBAF.

A list of the speakers and topics is given in Table~\ref{tab:talks}; the
remainder of this report summarizes and expands on their comments. For
additional details see the individual contributions to this report.

\section{Conventional Spectroscopy}

A variety of important problems in conventional $q\bar q$ meson spectroscopy
($q=u,d,s$) can be addressed at CEBAF using photon beams. Establishing $q\bar
q$ spectroscopy is important both because anomalous states must be identified
against a background of these conventional resonances (which are poorly known
above about 1${1\over 2}$ GeV), and because there may yet be surprises in the
conventional light meson spectrum, for example in configuration mixing,
photon
couplings or decay amplitudes. The production mechanism at CEBAF may be
especially useful for distinguishing conventional $q\bar q$ mesons from
unusual
states, because their electroproduction amplitudes should be quite
characteristic and can be studied using polarized and $Q^2<0 $ virtual
photons.

One example of an interesting area in conventional mesons, $s\bar s$
spectroscopy, was discussed by Steve Godfrey in his plenary talk. Strangeonia
are attractive in part because they are largely unexplored, and
photoproduction
at CEBAF should allow a detailed study of $s\bar s$ spectroscopy. The strange
quark is sufficiently light so that one would naively expect $s\bar s$
spectroscopy to repeat $I=0$ $(u\bar u + d\bar d)$ spectroscopy, displaced
upwards by $\approx 200$-$250$ MeV, but surprises such as $\eta-\eta'$ flavor
mixing may await experimental studies. The spectroscopy of radially-excited
$^1S_0$ $q\bar q$ states is especially interesting because of the possibility
of $\eta-\eta'$ type mixing and because several poorly understood $0^{-+}$
states reported in $\psi(3100)$ radiative decays, such as the $\eta(1440)$,
may
involve radially-excited $q\bar q$ states. The $0^{++}$ channel is
interesting
because the lightest glueball is expected to be a scalar with a mass of
$\approx 1.5$ GeV, just where the $^3P_0$ $s\bar s$ scalar is expected.
Strong
mixing between these and the nonstrange $I=0$ $^3P_0$ $q\bar q$ states may be
present, or these may be relatively unmixed states. A comparison of $s\bar s$
and $(u\bar u \pm d\bar d)$ spectroscopy may also be useful as an indication
of
the importance of mixing with two-meson continuua, since this effect is
expected to split the $(u\bar u \pm d\bar d)$ $I=0,1$ $^3D_1$ states near 1.6
GeV as well as the $(u\bar u \pm d\bar d)$ $^3S_1'$ radials near 1.45
GeV,\cite{PG} and may induce important $s\bar s$ components in the $I=0$
states. These predictions may be testable through the photoproduction
amplitudes and branching fractions of these states.

The best experimental work to date on $s\bar{s}$ spectroscopy has been done
by
the SLAC E135 collaboration\cite{E135_ssbar} using the reactions
$K^-P\rightarrow K^-K^+\Lambda$, $K_S^0K_S^0\Lambda$ and
$K_S^0K^\pm\pi^\mp\Lambda$.  Photoproduction would add considerably to our
knowledge since $s\bar{s}$ pairs are copiously produced.  However, many of
the
interesting $s\bar s$ states are expected to lie near or above
2~GeV;\cite{Godfrey_Isgur} the $L=2$ and $L=3$ $s\bar s$ multiplets are
expected at 1.9 and 2.2 GeV respectively, and the $L=0$ $s\bar s$ radials are
expected at 1.6-1.7 GeV. Producing states above 2 GeV would require higher
beam
energies than are currently available at CEBAF, particularly if virtual
photons
are used.

Simon Capstick discussed baryon spectroscopy, and noted that although a
considerable program of baryon spectroscopy is planned for CEBAF with
$E_\gamma\leq4$~GeV (especially relating to photo- and electroproduction
amplitudes), much more could be done with a higher beam energy. For example,
the $Q^2$-dependence of electroproduction amplitudes of problematical
resonances such as the Roper could be measured usefully
(from the quark model viewpoint) to $\approx 5$ Gev$^2$
for comparison with theoretical predictions
for $q^3$-radial, hybrid or other assignments for this and other
controversial
states. In addition, associated production of strange baryons with kaons
could
be useful for the study of strange baryon spectroscopy, which is much less
well
known than nonstrange spectroscopy. Here interesting topics include the
$\Lambda(1405)$ (which is too light to be a conventional $uds$
baryon\cite{CI})
and the possible existence of $Z^*$ flavor-exotic molecules.\cite{Zstar} Only
moderate beam energies are required for strange baryon spectroscopy; to reach
the photoproduction threshold of $K+\Lambda(1405)$ we require only
$E_\gamma=1.5$~GeV, and photoproduction of a $\bar K$ and a 2.1~GeV $Z^*$
requires $E_\gamma=3.1$~GeV.

Nimai Mukhopadhyay discussed new calculations of baryon resonance production
in
the $\gamma P\rightarrow\eta P$ and $\gamma P\rightarrow\eta^\prime P$
reactions.  The calculations indicate that the cross section for
$\eta^\prime$
production, at least near threshold, depends on $s$-channel creation of the
$N^*(2080)$.  Again, the $Q^2$ dependences of such reactions may provide new
clues to baryon structure.

Stan Brodsky urged us to consider reactions which would produce charmonium,
for
example $\gamma P\rightarrow\eta_c P$ and $\gamma P\rightarrow\psi P$, to
investigate the possibility that charmonium may bind to nuclei.\cite{SB}
Since
charmonium in flight may quickly dissociate in nuclei, open charm reactions
such as $\gamma P\rightarrow \bar D^0\Lambda_c^+$ should also be studied. The
threshold photon energies for these processes are rather high, $E_\gamma =
7.7$~GeV for $\eta_c P$, $E_\gamma = 8.2$~GeV for $\psi P$ and $E_\gamma =
8.7$~GeV for $\bar D\Lambda_c$.

\section{Exotica: Molecules and Hybrid Mesons}

A primary goal of modern meson spectroscopy is the identification of
``exotica", which are mesons external to the $q\bar q$ quark model. The more
restrictive term {\it exotic meson} is used to refer to a meson whose quantum
numbers are inconsistent with {\em any} $q\bar{q}$ assignment, either because
of flavor (for example $I=2$)
or $J^{PC_n}$;
$0^{--}$,
$0^{+-}$,
$1^{-+}$,
$2^{+-}$,
$3^{-+}$\dots are all forbidden to $q\bar q$ states.
Exotica includes multiquark systems (such as weakly bound hadronic molecules)
and states with gluonic excitation, generically called gluonic hadrons.
Gluonic
hadrons are referred to as hybrids if quarks and gluonic excitations are both
present in the dominant basis state, and glueballs if they are relatively
pure
excited glue.

Much of the recent interest in multiquark systems has concentrated on
molecules, which are weakly bound states of two or more conventional $q\bar
q$
or $qqq$ hadrons. Nuclei are the most familiar examples of these. Two meson
resonances, the $f_0(975)$ and $a_0(980)$ have long been advocated by
Weinstein
and Isgur\cite{WI} as $K\bar K$ molecules. These resonances are a problem in
the quark model if one tries to identify them as $^3P_0$ $q\bar q$ states,
due
to their low masses and anomalous strong and electromagnetic couplings. In
our
working group John Weinstein presented the theoretical case for identifying
these states as $K\bar{K}$ molecules, and Alex Dzierba discussed an
experiment\cite{Phirad} at CEBAF which can distinguish between $K\bar K$ and
$q\bar q$ assignments by measuring the branching ratios\cite{CKI} for
$\phi\rightarrow\gamma f_0(975)$ and $\phi\rightarrow\gamma a_0(980)$. There
are many other candidates for light molecules,\cite{othermol} notably the
$\Lambda(1405)$ baryon ($\bar KN$) and the mesons $f_1(1420)$ ($K^*\bar K +
h.c.$ enhancement), $f_0(1520)$ ($\rho\rho , \omega\omega $) and $f_0(1710)$
($K^*\bar K^* , \omega\phi $), which CEBAF might also usefully study in
photoproduction experiments.

Curtis Meyer summarized results from the Crystal Barrel $P\bar P$
experiment\cite{CBar} at LEAR, including high-statistics results on the
$f_0(1520)$ in $\pi^o\pi^o$ and $\eta\eta$. This state is interesting as a
candidate ($\rho\rho , \omega\omega $) molecule and as a candidate scalar
glueball. (A search for flavor-singlet couplings, indicated by the $\pi\pi$,
$K\bar K$ and $\eta\eta$ branching fractions, may tell us if this state is
indeed the anticipated 1.5 GeV scalar glueball; the $K\bar K$ analysis is in
progress.) Meyer's talk emphasized the importance of correctly including
interference effects and unitarity in extracting broad, overlapping
resonances
in the 1-2 GeV mass range.

Suh-Urk Chung discussed the status of $J^{PC_n}$-exotics, notably hybrids in
the flux-tube model, and summarized current experiments and future
experimental
prospects. $J^{PC_n}$-exotic hybrids arise when the excited gluonic degree of
freedom is combined with the $q\bar q$ system to make a hybrid meson with
definite quantum numbers; these exotic states are predicted by all the
theoretical models and approaches which have been applied to
hybrids.\cite{hyrevs} The quantum numbers $1^{-+}$ are often chosen for
experimental searches because most models predict this to be the lightest
exotic hybrid, and the flux tube model of decays predicts the $I=1$ state to
be
relatively narrow.

Much of the current interest in hybrids is due to the flux tube model of
Isgur,
Kokoski, Merlin and Paton.\cite{IsgurExotic} This model treats the confining
$q\bar{q}$ interaction as a gluonic flux tube between the quarks, and makes
quite specific predictions for the masses and partial widths of states with
excited flux tubes, which are the hybrids of this model. The mass of the
lightest hybrid multiplet is predicted to be $\approx 1.9$~GeV, and the
anticipated dominant decay modes are quite characteristic. In the flux tube
model a hybrid meson decays by $q\bar q$ pair production, which breaks the
flux
tube. Consideration of overlap integrals suggests that the orbital angular
momentum of the flux tube in a decaying hybrid tends to be transferred to an
internal $L$ of a final $q\bar q$ meson, so the preferred decay modes of the
lowest-lying hybrids are the little-explored S+P two-body systems. The
constituent gluon model of hybrids also predicted this preferred
mode.\cite{conglu}

The low-lying exotic hybrids and their decays predicted by Isgur, Kokoski and
Paton are listed in Table~\ref{tab:exotics}.  The notation has been changed
to
conform to 1992 Particle Data Group useage.\cite{PDG92} In the table,
$\Gamma_{H\to AB}$ is the partial width to the channel specified, which
consists of a $\pi$ or a $K$ plus an excited meson.  (For the strange
particle
decays, our notation implies the two charge conjugate combinations that give
zero strangeness.) We also list the principal decay branches of the excited
meson ``B" in the two-body final state; $\Gamma_B$ is the total width of the
excited meson, and in the list of final states we abbreviate $K^*(892)$ by
$K^*$ and $K_0^*(1430)$ by $K_0^*$.

Chung noted that a persistent but controversial $I=1, 1^{-+}$ signal has been
reported in $\eta\pi$ and $\eta^\prime\pi$ by several
experiments.\cite{etapi}
He also presented new results from BNL\cite{E818} which appear consistent
with
a $1^{-+}$ exotic $\hat{\rho}(1900)$ decaying to $f_1(1285)\pi^-$ followed by
\mbox{$f_1\rightarrow K^+K_S^0\pi^-$}; an experiment in progress\cite{E852}
will study this candidate exotic state with better statistics.

The flux tube model suggests that photoproduction should be an effective way
to
produce hybrids, since the transfer of momentum and angular momentum to a
single struck quark may be an effective way to ``pluck" (orbitally excite) a
flux tube. There is some evidence for a $1^{-+}$ exotic, possibly a hybrid,
near 1.8-1.9 GeV in photoproduction of $b_1\pi$ (Atkinson {\it et
al.}\cite{Omega}) and $\rho\pi$ (Condo {\it et al.}\cite{Condo}) final
states.
Gary Adams discussed a proposal to study some of these final states at CEBAF
using the CLAS spectrometer in Hall~B.

In summary, there is an excellent opportunity for CEBAF to contribute to the
spectroscopy of exotica through photoproduction studies if the beam energy is
increased to $\approx 6$~GeV or beyond.

\section{$CP$ and $CPT$ Violation in $\phi$ Decay}

One very exciting possibility for CEBAF is the study of $CP$ and $CPT$
violation in $\phi$ decay.  This idea was originally developed\cite{UCLA} to
take advantage of $e^+e^-$ $\phi$-factories such as those now under
construction at Frascati and Novosibirsk. In these experiments one measures
the
decays $K^0\rightarrow\pi^+\pi^-$ and $K^0\rightarrow\pi^0\pi^0$ in the same
apparatus (to reduce systematic errors) and determines the asymmetry between
these two modes as a function of the separation of the $K_L^0$ and $K_S^0$
from
the $\phi$ decay. Not only is this an intriguing way to measure
$\epsilon^\prime/\epsilon$ from $K^0$ decay, it also provides a new test of
$CPT$ violation. This may be more than an exercise in improving lower limits,
since some string theories predict $CPT$ violation.\cite{AlanK}

To carry out these measurements at CEBAF, the plan is to produce ``flying''
$\phi$ mesons using the reaction \mbox{$\gamma P\rightarrow\phi P$}.  This
was
the principal topic of a recent workshop at Indiana University; Alex Dzierba
summarized the conclusions of that meeting and the status of this work in his
plenary talk. CEBAF has a {\em potential} advantage in that it generates a
much
larger $\phi$ luminosity than the $e^+e^-$ machines, although the associated
problems of detector design and especially background rejection are
nontrivial.
Since the decay products are boosted forwards, particle identification is
considerably easier than at an $e^+e^-$-$\phi$ factory.

A possible problem with a $CPT$ experiment at CEBAF is the presence of a
substantial background from S-wave $K\bar{K}$ photoproduction.\cite{Kswave}
This introduces a $K_S^0K_S^0$ signal which could obscure the $CP$-violating
amplitude from the decay of a $K_L^0$ from the $\phi$. However, Nathan Isgur
explained how the dominant S-wave can be used to advantage by observing the
interference between the S-wave and P-wave ($\phi$) amplitude.  These ideas
are
still under development, but appear promising at present.

\section{New Detectors}

Our working group also discussed possible new detector systems that could be
used for both spectroscopy and $\phi$ decay experiments.  Of the existing
CEBAF
detectors, only CLAS is appropriate for these studies, and Gary Adams
discussed
possible exotic meson signatures that could be measured in the CLAS. It
appears
unlikely however that this general purpose facility will prove suitable for
the
complete range of experiments we considered.

In his plenary talk Alex Dzierba discussed an experiment\cite{Phirad} using a
new segmented lead glass calorimeter\cite{P2} to study $\phi$ radiative
decays.
This can be carried out on a relatively short time scale at CEBAF because the
detector is currently available. Although this experiment is limited to
all-neutral decay modes of the $\phi$, produced with a real, tagged photon
beam, it illustrates how impressive physics results may be forthcoming from a
relatively simple apparatus.

Experience with facilities such as LEAR have made clear the importance of
complete event reconstruction, one should have charged particle detection and
momentum measurement in addition to neutrals.  Since many results will
require
relatively detailed partial wave analyses, it is important that a detector
for
spectroscopy approach $4\pi$ solid angle coverage with as little bias as
possible. Experiments which would use an incident electron beam (as opposed
to
photons, in which case the primary electrons have been removed) face
additional
complications from luminosity limitations.

The traditional approach to high energy spectroscopy has been to use a dipole
magnet spectrometer with a large magnetic field volume, equipped with
tracking
chambers, particle identification, and possibly neutral particle
detection.\cite{E852}  This approach has been applied to tagged photon
spectroscopy with good results,\cite{Omega,LAMP,OMEGA,SAPHIR} but in all
cases
the maximum allowed incident photon rate was limited to a
few$\times10^5$/sec.
This limitation arises partly from $e^+e^-$ pairs produced in the target by
the
incident photon beam; these pairs spread in the dipole field and are incident
on the detectors at very high rates.

At CEBAF, the CLAS avoids this problem, while retaining the large bending
power
of a dipole spectrometer, by using a toroidal magnetic field. This
essentially
eliminates the problem of $e^+e^-$ pair production, although as a result of
this geometry the detector is insensitive in the forward region because the
magnetic field is very low. Unfortunately, many of the high energy
experiments
of interest for spectroscopy, especially those using small-$Q^2$
``diffractive'' photon interactions, preferentially populate the
extreme-forward region.

An alternative approach, followed by Crystal
Barrel and LASS,\cite{Xbarrel,LASS} is to use a solenoidal magnetic field.
Since
particles are deflected only according to their transverse momentum, one can
sacrifice some momentum resolution for the more forward-going particles, and
LASS\cite{LASS} augment their system with a dipole magnet far downstream.
Advantages of solenoidal spectrometers include a large, flat acceptance over
phase space, and the ``trapping'' of $e^+e^-$ pairs along the axis of the
solenoid.  In our working group Dan Coffman discussed the state-of-the-art in
solenoidal spectrometers, specifically the CLEO detector at Cornell/CESR
which
is used for high energy $e^+e^-$ collision experiments.

Heavy quark spectroscopy, including charm and $\tau$ decays, would be
difficult to
pursue unless CEBAF increases its energy to well above 8~GeV.  (See Figure~1.)
Kam Seth proposed a different solution, namely an $e^+e^-$ $\tau$/charm
factory
using a 6~GeV electron beam incident on a 1~GeV positron beam in a storage
ring.  Center-of-mass energies up to 4.90~GeV are possible, and that would be
extended to 5.66~GeV with an 8~GeV electron beam.  What's more, the incident
electron beam is used ``non-destructively'', and might be incident on fixed
targets downstream.  Similar designs have been studied previously, and
luminosities approaching $10^{33}/$cm$^2$sec may be possible, leading
to orders of magnitude improvement in the present event sample.

Specific designs and cost estimates for these detector and
accelerator developments are in preparation.

\vfill\eject

\begin{table}
\begin{center}
\caption{Agenda for the Hadron Spectroscopy Working Group}
\label{tab:talks}
\begin{tabular}{ll}
\ \ \ & \ \ \ \\
\ \ \ Speaker & \ \ \ Topic\\
\hline\\
\multicolumn{2}{c}{\it Thursday 14 April}\\
\\
Simon Capstick (FSU) & Problems in Baryon Spectroscopy\\
Alex Dzierba$^*$ (IU) & Flying $\phi$s and High Energy Spectroscopy\\
John Weinstein (U.Miss.) & $K\bar{K}$ Molecules\\
\\
\hline\\
\multicolumn{2}{c}{\it Friday 15 April}\\
\\
Steve Godfrey$^*$ (Carleton) & Issues in Light Hadron Spectroscopy\\
Suh-Urk Chung (BNL) & Exotic Meson Spectroscopy\\
Curtis Meyer (CMU) & Results from Crystal Barrel\\
Kamal Seth (Northwestern) & A $\tau$-Charm Factory at CEBAF\\
Gary Adams (Rensselaer) & Exotic Meson Spectroscopy in CLAS\\
Dan Coffman (Cornell) & Solenoidal Detectors in Spectroscopy\\
\\
\hline\\
\multicolumn{2}{c}{\it Saturday 16 April}\\
\\
Marc Sher (William and Mary) & Gluino Searches at CEBAF Energies\\
Nimai Mukhopadhyay (Rensselaer) & Photoproduction of $\eta$ and
$\eta^\prime$\\
Nathan Isgur (CEBAF) & S-P Interference in $K\bar{K}$ Photoproduction\\
                     & ~~~and Searches for $CP$ and $CPT$ Violation\\
\multicolumn{1}{c}{$^*$ Plenary Talk} & \\
\hline
\end{tabular}
\end{center}
\end{table}

\begin{table}
\begin{center}
\caption{$J^{PC_n}$-Exotic Hybrid Mesons and Dominant
Decays in the Flux Tube Model}
\label{tab:exotics}
\begin{tabular}{|rrrlr|rlr|}
\multicolumn{5}{c}{\bf Hybrid Meson} & \multicolumn{3}{c}{\bf Secondary
Decays
}\\
\hline
&\multicolumn{1}{c}{IKP} &     &    &
$\Gamma_{H\to AB}$ & $\Gamma_B$ & &\\
State & label & $J^{PC_n}(I^G)$ & Decay & (MeV) &
(MeV) & Decay & $b.f.$\\
\hline
$\hat{a}_2(1900)$ & $x_2^{+-}$ & $2^{+-}(1^+)$ &
$\left[\pi a_2(1320)\right]_P$ & 450 &
103 & $a_2\rightarrow\rho\pi$ & 70\%\\
& & & & & & $\rightarrow\eta\pi$ & 15\%\\
& & & & & & $\rightarrow\omega\pi\pi$ & 11\%\\
& & & & & & $\rightarrow K\bar{K}$ & 5\%\\
& & & $\left[\pi a_1(1260)\right]_P$ & 100 &
$\sim$400 & $a_1\rightarrow\rho\pi$ & most\\
& & & $\left[\pi h_1(1170)\right]_P$ & 150 &
360 & $h_1\rightarrow\rho\pi$ & seen\\
\hline
$\hat{f}_2(1900)$ & $y_2^{+-}$ & $2^{+-}(0^-)$ &
$\left[\pi b_1(1235)\right]_P$ & 500 &
155 & $b_1\rightarrow (\omega\pi)_{S,D}$ & most\\
& & & & & & $\rightarrow \eta\rho$ & seen\\
\hline
$\hat{f}_2^\prime(2100)$ & $z_2^{+-}$ & $2^{+-}(0^-)$ &
$\left[K K_2^*(1430)\right]_P$ & 250 &
 98 & $K_2^*\rightarrow K\pi$ & 50\%\\
& & & & & & $\rightarrow K^*\pi$ & 25\%\\
& & & & & & $\rightarrow K^*\pi\pi$ & 13\%\\
& & & $\left[KK_1(1400)\right]_P$ & 200 &
174 & $K_1\rightarrow K^*\pi$ & 94\%\\
\hline
$\hat{\rho}(1900)$ & $x_1^{-+}$ & $1^{-+}(1^-)$ &
$\left[\pi b_1(1235)\right]_{S,D}$ & 100,30 &
155 & $b_1\rightarrow (\omega\pi)_{S,D}$ & most\\
& & & & & & $\rightarrow \eta\rho$ & seen\\
& & & $\left[\pi f_1(1285)\right]_{S,D}$ & 30,20 &
 24 & $f_1\rightarrow\eta\pi\pi$ & 50\%\\
& & & & & & $\rightarrow4\pi(\rho\pi\pi)$ & 38\%\\
& & & & & & $\rightarrow a_0(980)\pi$ & 37\%\\
\hline
$\hat{\omega}(1900)$ & $y_1^{-+}$ & $1^{-+}(0^+)$ &
$\left[\pi a_1(1260)\right]_{S,D}$ & 100,70 &
$\sim$400 & $a_1\rightarrow\rho\pi$ & most\\
& & &
$\left[\pi\pi(1300)\right]_P$ & 100 &
200-600 & $\pi(1300)\rightarrow\rho\pi$ & seen\\
& & &
$\left[KK_1(1400)\right]_S$ & 100 &
174 & $K_1\rightarrow K^*\pi$ & 94\%\\
\hline
$\hat{\phi}(2100)$ & $z_1^{-+}$ & $1^{-+}(0^+)$ &
$\left[KK_1(1270)\right]_D$ & 80 &
90 & $K_1\rightarrow K\rho$ & 42\%\\
& & & & & & $\rightarrow K_0^*\pi$ & 28\%\\
& & & & & & $\rightarrow K^*\pi$ & 16\%\\
& & & & & & $\rightarrow K\omega$ & 11\%\\
& & & $\left[KK_1(1400)\right]_S$ & 250 &
174 & $K_1\rightarrow K^*\pi$ & 94\%\\
& & & $\left[KK(1460)\right]_P$ & 30 &
250 & $K\rightarrow K\pi\pi$ & seen \\
\hline
$\hat{a}_0(1900)$ & $x_0^{+-}$ & $0^{+-}(1^+)$ &
$\left[\pi a_1(1260)\right]_P$ & 800 &
$\sim$400 & $a_1\rightarrow\rho\pi$ & most\\
& & & $\left[\pi h_1(1170)\right]_P$ & 100 &
360 & $h_1\rightarrow\rho\pi$ & seen\\
& & & $\left[\pi\pi(1300)\right]_S$ & 900 &
200-600 & $\pi(1300)\rightarrow\rho\pi$ & seen\\
\hline
$\hat{f}_0(1900)$ & $y_0^{+-}$ & $0^{+-}(0^-)$ &
$\left[\pi b_1(1235)\right]_P$ & 250 &
155 & $b_1\rightarrow (\omega\pi)_{S,D}$ & most\\
& & & & & & $\rightarrow \eta\rho$ & seen\\
\hline
$\hat{f}_0^\prime(2100)$ & $z_0^{+-}$ & $0^{+-}(0^-)$ &
$\left[KK_1(1270)\right]_P$ & 800 &
90& See $\hat{\phi}(2100)$ &\\
& & &  $\left[KK_1(1400)\right]_P$ & 50 &
174 & $K_1\rightarrow K^*\pi$ & 94\%\\
& & & $\left[KK(1460)\right]_S$ & 800 &
250 & $K\rightarrow K\pi\pi$ & seen \\
\hline
\end{tabular}
\end{center}
\end{table}

\end{document}